\documentclass[universe,review,accept,oneauthor,pdftex,10pt,a4paper]{mdpi}
\firstpage{1} 
\articlenumber{4}
\doinum{10.3390/universe2010004}
\pubvolume{2}
\pubyear{2016}
\copyrightyear{2016}
\externaleditor{Academic Editors: Kazuharu Bamba and Sergei D. Odintsov}
\history{Received: 30 December 2015; Accepted: 3 February 2016; Published: 23 February 2016}
\pdfoutput=1
 \usepackage{amssymb}
 \usepackage{amsmath}
  \usepackage{amsfonts}
  \usepackage{epsfig}
   \usepackage{soul}
   \usepackage{bm} 
 \theoremstyle{mdpi}
 \newcounter{thm}
 \newcounter{ex}
 \newcounter{re}
 \theoremstyle{mdpidefinition}

\Title{Quantum Yang--Mills Dark Energy}

\Author{Roman Pasechnik}
\address[1]{%
 Department of Astronomy and Theoretical Physics, Lund
University, SE-223 62 Lund, Sweden; Roman.Pasechnik@thep.lu.se}

\corres{Correspondence: Roman.Pasechnik@thep.lu.se}

\abstract{In this short review, I discuss basic qualitative characteristics of quantum non-Abelian
gauge dynamics in the non-stationary background of the expanding Universe in the framework of the standard
Einstein--Yang--Mills formulation. A brief outlook of existing studies of cosmological Yang--Mills fields and
their properties will be given. Quantum effects have a profound impact on the gauge field-driven cosmological evolution. 
In particular, a dynamical formation of the spatially-homogeneous and
isotropic gauge field condensate may be responsible for both early and late-time acceleration, as well as
for dynamical compensation of non-perturbative quantum vacua contributions to the ground state of the Universe.
The main properties of such a condensate in the effective QCD
 theory at the flat Friedmann--Lema\'itre--Robertson--Walker (FLRW) background
will be discussed within and beyond perturbation theory. Finally, a phenomenologically consistent
dark energy can be induced dynamically as a remnant of the QCD vacua compensation arising from
leading-order graviton-mediated corrections to the QCD ground state.}



\begin{document}

\section{Introduction}

The cosmological constant (or, in general, dark energy (DE)) problem is one of the most controversial
and debatable naturalness problems in theoretical physics and cosmology nowadays. It refers to an enigmatic
(vacuum-like) anti-gravitating substance, which causes the Universe to expand with acceleration
typical for de-Sitter cosmologies (the late-time acceleration). The standard cosmological model known as the cold dark matter (CDM)
with the time-independent DE density is called the $\Lambda$-term (or $\Lambda$CDM) and agrees well with
the bulk of observational data, e.g., in studies of the Type Ia supernovae \cite{SNE1A-1,SNE1A-2}, cosmic microwave background
anisotropies \cite{WMAP-11,WMAP-12,WMAP-21,WMAP-22}, large-scale structure \cite{SDSS-1,SDSS-2}, {\it etc.} The recent Planck data \cite{Planck-1,Planck-2}
have further supported the cosmological constant (CC) hypothesis. While being so successful in observational cosmology,
the physical nature of the $\Lambda$-term with the vacuum equation of state consistent with phenomenology has not been
theoretically well understood yet and remains one of the major unsolved problems of theoretical physics
\cite{Weinberg:1988cp,Wilczek:1983as} (for recent reviews on this topic, see, e.g.,~\cite{our-review,Sola:2013gha}
and the references therein). On the way of searching for possible solutions of the DE problem, many various pathways were explored
during the past few decades typically referring to new exotic forms of matter. For a comprehensive review of existing
theoretical models and interpretations of the cosmological constant (or slowly-evolving DE), see, e.g.,~\cite{Peebles:2002gy,Padmanabhan:2002ji,Copeland:2006wr,Bamba:2012cp,Li:2012dt,Yoo:2012ug,Martin:2012bt} and the references
therein. Given such a huge variety of DE models in the literature, there is an apparent deficit of phenomenological data
capable of robustly constraining the possible time dependence of the DE density.

While observers are comfortable with taking the $\Lambda$-term as an additional parameter, which considerably
improves the $\Lambda$CDM fits to the data, for theorists, the uniform positive and small $\Lambda$-term raises
plenty of issues and contradictions with the existing quantum field theory (QFT). Many of the existing DE/CC models
cannot provide a natural explanation, not only of why $\Lambda$ is small and positive (the ``old'' CC problem), but also
why $\Lambda$ is non-zeroth and exists at all (the ``new'' CC problem). Clearly, a theoretically-consistent framework
should address both problems simultaneously. Ideally, such a framework should explain the smallness of the observed
$\Lambda$-term, either in terms of known fundamental constants or via a dynamical vacuum self-tuning mechanism.
Starting from a renormalizable field theory, one could simply fix an arbitrary $\Lambda$-term value at an arbitrary energy
scale. The basic problem, however, is to describe various quantum vacuum (condensate) contributions to the ground state
energy at macroscopic separations (IR limit), as well as their renormalization group (RG) running without having a complete high-energy
QFT (UV limit). Of course, the latter would consistently unify all four different types of interactions in Nature,
providing a naturally small positive $\Lambda$-term, as well as containing a suitable candidate for the inflaton field.
However, certain aspects of the early/late time acceleration could be, in principle, addressed even before such a theory
has been developed. One of the compelling directions theorists take in order to address the CC problem from the first
principles is on the way towards a better understanding of quantum dynamics of the ground state of the Universe and
its evolution in time, as well as its possible relation to the late time acceleration.

Various non-Abelian fields are commonly present in the Standard Model (SM)
 and its high-energy extensions, playing an important role in both particle physics
and cosmology, along with scalar (e.g., Higgs) fields. For a detailed discussion of the various cosmological implications of gauge fields, we refer
to a vast literature on the subject, e.g.,~\cite{Galtsov:1991un,Cavaglia:1993en,Bamba:2008xa,Elizalde:2010xq,Galtsov:2011aa,
Maleknejad:2011sq,Elizalde:2012yk,Rinaldi:2015iza}. In particular, such a strongly-coupled system as the Bose--Einstein condensate
of gluons in quantum chromodynamics (QCD) is responsible for spontaneous chiral symmetry breaking, as well as for the color confinement
(for a comprehensive review on the topological QCD vacuum, see, e.g.,~\cite{Shifman-1,Shifman-2,Schafer:1996wv,Diakonov:2002fq,Diakonov:2009jq}
and the references therein). Does a physical mechanism or a dynamical principle exist within the conventional QFT and particle physics framework that
can be responsible for the late-time (DE/CC-driven) acceleration? This review aims at the search for a comprehensive answer to this fundamental
question within the framework of strongly-coupled quantum Yang--Mills (YM) field theories with a non-trivial ground state, such as~QCD.

\section{Vacuum Catastrophe}

One typically introduces the CC into the classical action of the gravitational field,
such that the total action accounting for both gravity and matter fields (with spin less than two,
e.g., scalar $\phi$, spinor $\psi$ and vector $A_\mu$ fields), as well as their interactions with each
other and with external gravitational field $g_{\mu\nu}$ reads:
\begin{eqnarray}
S=-\int d^4x \sqrt{-g} \Big(\frac{R}{2\kappa}+\Lambda_0\Big)+S_m[\phi,\psi,A_{\mu},g_{\mu\nu}] \,, \qquad \Lambda_0 = \rm{const} \,,
\end{eqnarray}
where $\kappa=8\pi G$ in terms of the gravitational constant $G=M_{\rm PL}^{-2}$ and
the Planck mass $M_{\rm PL}=1.2\cdot 10^{19}$ GeV. The resulting Einstein equations of motion
for the macroscopic geometry:
\begin{eqnarray}
R_{\mu\nu} - \frac{1}{2} g_{\mu\nu}R=\kappa(\Lambda_0 g_{\mu\nu} + T_{\mu\nu})\,, \qquad
T_{\mu\nu} = - \frac{2}{\sqrt{-g}}\frac{\delta S_m}{\delta g^{\mu\nu}} \,,
\end{eqnarray}
should be accompanied by the respective equations of motion for the matter fields.

So far, the $\Lambda_0$-term is just an arbitrary constant parameter allowed by Lorentz invariance, whose value and sign
cannot be predicted within the classical field theory alone. This situation considerably changes in QFT
possessing a non-trivial ground state, such that the averaged energy-momentum tensor of all of the fields
present in the Universe $T_{\mu\nu}$ over the Heisenberg vacuum state:
\begin{eqnarray}
\langle 0 |T_{\mu\nu}|0\rangle = \Lambda_{\rm vac}(\mu) g_{\mu\nu}\,, \qquad \Lambda_{\rm vac}(\mu) \not=0
\end{eqnarray}
yields a non-trivial energy density of the {quantum vacuum} \cite{Sakharov-1,Sakharov-2}, which depends
on the renormalization scale $\mu$ and satisfies the vacuum equation of state:
\begin{eqnarray}
p_{\Lambda} = w \Lambda_{\rm vac} \,, \qquad w=-1 \,.
\end{eqnarray}
The arbitrariness in the classical (or ``bare'') contribution $\Lambda_0$
is to be eliminated by a measurement at a fixed scale $\mu=\mu_{IR}$ corresponding to the present Universe, \textit{i.e}.,
\begin{eqnarray} \label{vac-eos}
\Lambda_{\rm cosm}\equiv \Lambda_0+\Lambda_{\rm vac}(\mu_{IR})\,.
\end{eqnarray}
In standard QFT, one often ignores the vacuum energy; it serves as a reference point, while one is interested
in microscopic properties, such as the masses and energies of the excitations about the vacuum state. In general relativity (GR), however,
the vacuum energy naturally gravitates (if one adopts silently that gravitational iterations are fundamental) and, thus, affects
the cosmological evolution. Obviously, the quantum vacuum contributions cannot be eliminated independently at every
distinct energy scale by, e.g., a naive shift of the zeroth level of the vacuum energy. The ground state of the Universe
therefore accounts for the whole bulk of various contributions from existing quantum fields at energy scales
ranging from the quantum gravity (Planck) scale, $M_{\rm PL}\sim 10^{19}$ GeV, down to the QCD confinement scale,
$M_{\rm QCD} \sim 1$ GeV. These are the well-known maximal and minimal energy scales of particle physics, respectively; the former
determines the strength of gravitational interactions, while the latter sets the characteristic time scale of the last QCD
phase transition at which the current ground state of the Universe has been created. An accurate analysis of various
contributions to the ground-state energy of the Universe accounting for typical zero-point fluctuations of boson and
fermion fields, as well as the non-trivial minimum of the classical Higgs field has been done in~\cite{Martin:2012bt,Sola:2013gha}.

In what follows, we distinguish between the weakly-coupled (perturbative) and strongly-coupled (non-perturbative) vacua.
In particle physics, among the well-known vacuum subsystems of the standard model are the Higgs condensate (conventionally, weakly-coupled classical
subsystem), responsible for the spontaneous electroweak symmetry breaking in the SM, and the quark-gluon condensate (strongly-coupled quantum subsystem),
responsible for the spontaneous chiral symmetry breaking and color confinement in QCD. By a very rough estimate, the electroweak symmetry breaking scale provided
by the vacuum expectation value of the Higgs field, $H(x)=\langle 0 | H | 0 \rangle + h(x)$, $\langle 0 | H | 0 \rangle \sim 100$ GeV, gives rise to
the Higgs condensate contribution:
\begin{eqnarray}
\label{Lambda-EW}
\Lambda_{\rm vac}^{\rm EW}\sim \langle 0 | H | 0 \rangle^4 \sim 10^8\, {\rm GeV}^4 \,.
\end{eqnarray}
The Higgs condensate is a classical homogeneous and isotropic component of the quantum Higgs field $H(x)$, which
determines the mass scale of weak gauge bosons $W^\pm$ and $Z^0$, as well as fermions in the SM.

The ground state in QCD is determined by non-vanishing quantum condensates of strongly interacting quarks and gluons typically
referred to as the quark-gluon condensate. This vacuum subsystem has unique properties and is responsible for the
confined phase of quark matter. According to one of the popular interpretations of the QCD vacuum, the topological (or instanton) modes
of the quark-gluon condensate are given by non-perturbative fluctuations of the gluon and sea (mostly, light) quark fields induced
in the processes of quantum tunneling of the gluon vacuum between topologically different classical states \cite{Shifman-1,Shifman-2,Schafer:1996wv,Diakonov:2002fq,
Diakonov:2009jq}. The topological contribution to the energy density of the QCD vacuum can be derived from the well-known trace anomaly
relation \cite{trace-anomaly-1,trace-anomaly-2,trace-anomaly-3}:
\begin{equation}
\label{traceA}
T_{\mu}^{\mu,{\rm QCD}}=\frac{\beta(g_s^2)}{2}F_{\mu\nu}^aF^{\mu\nu}_a+\sum_{q=u,d,s}m_q\bar{q}q\,,
\end{equation}
where $T_{\mu}^{\mu,{\rm QCD}}$ is the trace of the energy-momentum tensor of QCD, $g_s$ and $\beta(g_s^2)$ are 
the QCD coupling constant and the $\beta$-function, respectively, $m_q$ are the light (sea)
quark masses, and $F_{\mu\nu}^a$ is the gluon field stress tensor in the standard normalization. The vacuum average of the trace $\langle
0|T_{\mu}^{\mu,{\rm QCD}}|0\rangle$
\begin{eqnarray}
\Lambda_{\rm vac}^{\rm QCD}&=&-\frac{9}{32}\langle0|:\frac{\alpha_s}{\pi}F^a_{\mu\nu}(x)
F_a^{\mu\nu}(x):|0\rangle + \frac14 \Big[\langle0 | :m_u\bar uu: | 0\rangle + \langle0 | :m_d\bar dd: |0\rangle \nonumber \\
&+&
\langle0|:m_s\bar ss:|0\rangle\Big] \simeq -(5\pm 1)\times 10^{-3}\; \text{GeV}^4\,, \qquad \alpha_s=\frac{g_s^2}{4\pi} \,,
 \label{Lambda-QCD}
\end{eqnarray}
is composed of gluon and light sea $u,d,s$ quark contributions. This is the saturated (maximal) value of the topological contribution to the
QCD vacuum energy density, while its physical spacetime evolution, as well as other possible contributions to it, are not known yet, such that the
non-perturbative long-range YM dynamics remains poorly understood.

According to the observations, the CC density is positive and close to the critical density of the Universe today,
\begin{equation}
\label{Lcosm}
\Lambda_{\rm cosm}\simeq 0.7 \rho_{\rm crit} \simeq 2.5 \times 10^{-47}\, {\rm GeV}^4 >0\,, \qquad \rho_{\rm crit}\equiv \frac{3H_0^2}{\kappa} \,.
\end{equation}
Provided that the DE equation of state constrained by the measurements $w=-1.13^{+0.13}_{-0.10}$ is consistent with the pure
vacuum case (\ref{vac-eos}), the quantum vacuum is often considered to be a natural candidate for the DE. However, comparing
Equations~(\ref{Lambda-EW}) and (\ref{Lambda-QCD}) to Equation~(\ref{Lcosm}), one immediately notices a serious problem: the quantum
vacua contributions to the ground state energy of the Universe, despite that they are very different from each other, individually
exceed the observed CC value by many orders of magnitude. One would need to tune the input ``bare'' $\Lambda_0$ parameter
to cancel the net vacuum term $\Lambda_{\rm vac}$ to a degree of many tens of decimal digits in order to reach the observed
smallness of the CC, which is critical for, e.g., the structure formation in the Universe. Even if such a cancellation is achieved at
a given scale, quantum corrections will destroy it. This problem is often referred to as to the ``vacuum catastrophe'',
the fundamental failure of quantum physics in describing the macroscopic dynamics of the Universe as a whole, at least
in its current formulation \cite{Martin:2012bt,Sola:2013gha}.

One of the possible resolutions of this problem relies on a mechanism of dynamical compensation of short-distance vacuum fluctuations, in particular
during the electroweak and QCD phase transition epochs. In this scheme, no matter how the observed cosmological constant is interpreted in the end,
such huge quantum vacua contributions existing at short spacetime separations must be first eliminated {dynamically and separately} at
every distinct energy scale with enormous precision \cite{Polchinski:2006gy} in order to avoid a major fine tuning of unknown vacua parameters.
On the other hand, such a vacua self-alignment effect, if it takes place, should be generic for both \linebreak weakly- and strongly-coupled vacua subsystems
and, thus, may be regarded as a new physical phenomenon \cite{Copeland:2006wr,Dolgov-1,Dolgov-2,Pasechnik:2013sga,Pasechnik:2013poa}, which emerges
from yet unknown non-perturbative dynamics of the ground state in QFT composed of many interacting components.

From this point of view, the current ground state of the Universe with a small finite and positive $\Lambda$-term should be formed during a sequence
of phase transitions at very early times and reached today's density soon after the last QCD phase transition. Possible mechanisms for compensation
of weakly-coupled perturbative contributions to the net vacuum energy density of the Universe from vacuum fluctuations of fundamental fermions and bosons
(e.g., zero-point fluctuations, Higgs condensates in the SM and beyond, graviton condensate, \textit{etc}.) typically refer to high-scale supersymmetric
grand unified theories, supergravity and superstring theories (see, e.g.,~\cite{Copeland:2006wr,Nobbenhuis:2004wn,Nobbenhuis:2006yf} and the references therein).
In most popular extensions of the SM yet consistent with laboratory measurements, however, the supersymmetry is assumed to be explicitly broken,
which destroys the respective cancellation of the corresponding vacua. A true yet unknown high-scale theory, such as quantum gravity, is normally
expected to address this issue in a consistent way.

Given that the perturbative vacua can be eliminated (or excluded from consideration) by the time the Universe reaches the QCD phase transition
temperature $T_{\rm QCD}\sim 0.1$ GeV, a cancellation of the strongly-coupled non-perturbative quark-gluon condensate requires a dynamical
understanding of the QCD vacuum in the expanding Universe. Then, the observed cosmological constant can, in principle, be associated with an uncompensated
remnant formed soon after the chiral symmetry breaking during the latest QCD phase transition epoch. In order to explore such a natural possibility, one should
study the dynamical properties of the spatially-homogeneous and isotropic YM condensates in the expanding Universe at both classical and quantum levels.

An alternative way of avoiding the ``vacuum catastrophe'' was proposed recently in~\cite{Kaloper:2013zca} and refers to ``sequestering'' of all microscopic
vacuum contributions of the matter sector (including quantum corrections) from gravity, which is possible in the Universe finite in spacetime and collapsing in the future.
In a similar spirit,~\cite{Thomas:2009uh} exploits the promising idea about gravitation being not a truly fundamental interaction, but rather a low energy effective
interaction, such that gravitons should be treated as quasiparticles, which do not ``feel'' all microscopic degrees of freedom up to the Planck energy, but rather, interact
with a few certain excitations only in the IR limit of a yet unknown fundamental theory. In this case, one naturally expects a zeroth ``renormalized'' CC in a Minkowski
vacuum where the Einstein equations are automatically satisfied, while the observed DE is naturally determined by the deviation from the Minkowski spacetime geometry
(see, e.g.,~\cite{Thomas:2009uh,Schutzhold:2002pr,Klinkhamer:2007pe,Maggiore:2010wr}). The scale of such an effective theory of gravity emerges due to
the conformal (trace) anomaly, which is naturally present in QCD \cite{trace-anomaly-1,trace-anomaly-2,trace-anomaly-3} providing a non-vanishing contribution
to the vacuum energy~\cite{Schutzhold:2002pr} (see also~\cite{Sola:2013gha} and the references therein):
\begin{eqnarray}
\label{effectiveG}
\langle T^{\mu}_{\mu} \rangle \sim H \Lambda_{\rm QCD}^3 \sim (10^{-3}\, {\rm eV})^4 \,.
\end{eqnarray}
The latter is remarkably close to the observed CC today. We will come back to this interesting effect, which can also be seen within the philosophy of vacua
compensation, at a more quantitative level~below.

The philosophy of effective gravity being not sensitive to microscopic quantum vacua fluctuations above a certain energy scale, which may justify the late-time acceleration
and the observed smallness of the CC, however, may be questioned by an early-time acceleration mechanism for which to work out the gravitational interactions should
be able to ``resolve'' the quantum fluctuations of the inflaton field at much higher energy scale than the modern CC. Needless to remind, the scale-invariant perturbations
spectrum is a specific prediction of the quasiclassical theory describing interactions of the inflaton field and metric fluctuations at characteristic energies far beyond
the electroweak scale. Therefore, the effective gravity approach justifying the troublesome ``insensitivity'' of gravity to microscopic vacua (such as the Higgs condensate)
may have difficulties in the interpretation of early-time acceleration. In one way or another, most of the existing analyses of the observed DE simply ignore the huge microscopic
vacua terms, although the non-perturbative QCD ground state has a special status and should be treated carefully.

Now, we turn to a description of YM dynamics in cosmology and start with the classical case.

\section{Yang--Mills Condensates in Cosmology}
\label{Sec:YM-Cosm}

The gauge-invariant Lagrangian of the classical YM field in the $SU(N)$
($N=2,3,\dots$) gauge theory reads:
\begin{equation}
\mathcal{L}_{\rm cl}=-\frac{1}{4}F^a_{\mu\nu}F_a^{\mu\nu}\,, \label{L}
\end{equation}
where:
\begin{equation*}
F^a_{\mu\nu}=\partial_\mu A^a_\nu - \partial_\nu A^a_\mu +
g_{\rm YM}\,e^{abc} A^b_\mu A^c_\nu
\end{equation*}
is the YM stress tensor with isotopic (adjoint rep) $a,b,c=1,\dots\,, N^2-1$ and Lorentz
$\mu,\nu=0,1,2,3$ indices. Here, $g_{\rm YM}$ is the gauge coupling constant.
The corresponding generating functional of such a theory is given by the Euclidean
functional integral:
\begin{eqnarray} \label{FuncI}
Z\propto \int [DA]\, e^{-S_{\rm cl}[A] + \int J^a_\mu A^a_\mu d^4x} \,, \qquad
S_{\rm cl}[A]=\int \mathcal{L}_{\rm cl} d^4x \,,
\end{eqnarray}
which is dominated by minima of the classical action $S_{\rm cl}[A]$, unaltered
by quantum corrections. Such minima correspond to the classical vacuum state
with $F^a_{\mu\nu}=0$, while finite gauge field excitations about the classical
vacuum are known as instantons \cite{Shifman-1,Shifman-2,Schafer:1996wv,Diakonov:2002fq,Diakonov:2009jq}.

For practical purposes, one typically employs the temporal (Hamilton) gauge in
which the asymptotic states of the $S$-matrix automatically contain the physical
transverse modes only. The corresponding gauge condition reads:
\begin{equation}
A^a_0=0\,. \label{gauge}
\end{equation}
In the $SU(2)$ gauge theory, due the local isomorphism of the isotopic $SU(2)$ gauge group
and the $SO(3)$ group of spatial three-rotations, the unique (up to a rescaling) $SU(2)$
YM configuration can be parameterized in terms of a scalar time-dependent
spatially-homogeneous field \cite{Cervero:1978db,Henneaux:1982vs,Hosotani:1984wj}.
Introducing a mixed space-isotopic orthonormal basis $e^a_i$, $a,i=1,2,3$ in the temporal gauge
(\ref{gauge}), such that the YM field $\mathcal{A}_\mu^a$ transforms into a tensor field
$\mathcal{A}_{ik}$ as follows:
\begin{equation}
\label{e-basis}
e^a_i A^a_k\equiv A_{ik}\,, \qquad e^a_i e^a_k=\delta_{ik}\,,\qquad e^a_i e^b_i=\delta_{ab}\,.
\end{equation}
Then, the resulting spacial tensor $\mathcal{A}_{ik}$ can be separated into two parts:
\begin{equation}
A_{ik}(t,\vec x)=\delta_{ik}V(t)+\widetilde{A}_{ik}(t,\vec x)\,, \qquad
\langle \widetilde{A}_{ik}(t,\vec x) \rangle = \int d^3x \widetilde{A}_{ik}(t,\vec x) = 0 \,,
\label{TheFirst}
\end{equation}
where the time-dependent function $V(t)$ is identified with the isotropic and homogeneous classical YM condensate,
and $\widetilde{A}_{ik}(t,\vec x)$ are the spatially-inhomogeneous YM wave modes. A homogeneous YM condensate
can also be extracted in extended gauge theories (e.g., $SU(N)$), whose gauge group contains at least one $SU(2)$ subgroup.
In the QFT formulation, the YM wave modes are interpreted as YM quanta (e.g., gluons), while $V(t)$
contributes to the ground state of the theory, which is thus nontrivial and has to be studied in detail.

The separation into spatially-homogeneous and -inhomogeneous components in Equation~(\ref{TheFirst}) is analogical to the
conventional QCD instanton theory where one performs a mapping of three-space onto $SU(2)$ subgroup elements of
the color $SU(3)_c$. Besides, the well-known 't Hooft--Polyakov monopole~\cite{tHooft,Polyakov} is introduced
by means of an antisymmetric matrix with mixed Lorentz-isotopic indices in the Hamilton gauge, while the extracted
YM condensate $V=V(t)$ provides a symmetric analogue of such a solution. In practice, there are not any physical
arguments that could forbid the existence of the homogeneous non-Abelian condensate with an isotropic
energy-momentum tensor~\cite{Cembranos:2012ng} originating from unbroken $SU(N)$ gauge symmetry
at cosmological scales.

After a suitable covariant generalization of the classical YM action (\ref{FuncI}), the Einstein--Yang--Mills (EYM) equations of
the classical $SU(2)$ theory read:
\begin{eqnarray}
&& \frac{1}{\kappa}\left(R_\mu^\nu-\frac12\delta_\mu^\nu
R\right)=\frac{1}{\sqrt{-g}}\frac{1}{g_{\rm YM}^2}\left(
-\mathcal{F}_{\mu\lambda}^a\mathcal{F}^{\nu\lambda}_a + \frac14\delta_\mu^\nu
\mathcal{F}_{\sigma\lambda}^a\mathcal{F}^{\sigma\lambda}_a\right)\,, \nonumber\\
&& \left(\frac{\delta^{ab}}{\sqrt{-g}}\partial_\nu\sqrt{-g}-e^{abc}\mathcal{A}_\nu^c\right)
\frac{\mathcal{F}_b^{\mu\nu}}{\sqrt{-g}}=0\,, \label{classicalYM}
\end{eqnarray}
where:
\begin{eqnarray}
\mathcal{A}_\mu^a\equiv g_{\rm YM}A_\mu^a \,, \qquad \mathcal{F}^a_{\mu\nu}\equiv g_{\rm YM}F^a_{\mu\nu}=
\partial_\mu\mathcal{A}^a_\nu-\partial_\nu\mathcal{A}^a_\mu
+f^{abc}\mathcal{A}_\mu^b\mathcal{A}_\nu^c \,, \qquad U(t)\equiv g_{\rm YM} V(t) \,. \label{AF}
\end{eqnarray}
In the spatially-flat Friedmann--Lema\'itre--Robertson--Walker (FLRW) conformal metric:
\[
g_{\mu\nu}=a^2(\eta)\,{\rm diag}(1,\,-1,\,-1,\,-1) \,, \qquad \sqrt{-g}=a^4(\eta)\,, \qquad t = \int a(\eta) d\eta
\]
in zeroth order in small YM wave modes $|U|\gg g_{\rm YM} |\widetilde{A}_{ik}(t,\vec x)|$,
Equation (\ref{classicalYM}) reduces to the equations of motion for the YM condensate
$U=U(\eta)$ and the cosmological expansion law $a=a(\eta)$:
\begin{eqnarray}
\frac{3}{\kappa}\frac{a'^2}{a^4}=\frac{3}{2g_{\rm YM}^2a^4}\,\Big(U'^2+U^4\Big)\,,
\qquad U''+2U^3=0\,,
\label{class}
\end{eqnarray}
Its general solution corresponds to non-linear oscillations \cite{Pasechnik:2013sga}:
\begin{eqnarray}
&& U'^2+U^4=C\,,\qquad \int_{U_0}^U\frac{dx}{\sqrt{C-x^4}}=\eta \,, \qquad U_0,\,C = {\rm const} \,, \\
&& U'(0)=0\,, \qquad U_0=C \quad \to \quad A(\eta)\simeq U_0\cos\biggl(\frac65\,U_0\eta\biggr) \,.
\label{classol}
\end{eqnarray}
Therefore, the classical YM condensate behaves as an ultra-relativistic medium with energy density
$\varepsilon_{\rm YM}\sim 1/a^4$ and equation of state $p_{\rm YM}=\varepsilon_{\rm YM}/3$
\cite{Galtsov:1991un,Cembranos:2012ng}. The YM condensates in $SU(N)$ gauge theories obey similar equations
of motion, which may differ by a rescaling of the coupling constant affecting the frequency of
YM condensate oscillations only. In addition, the classical YM condensate coupled exponentially to
a scalar field (with an exponential potential) provides interesting solutions for early-time acceleration
thoroughly discussed in~\cite{Maeda:2012eg}, while classical YM models of DE were discussed in,
e.g.,~\cite{ArmendarizPicon:2004pm,Kiselev:2004py,Wei:2006tn,Jimenez:2008au} (for more
details, see, e.g.,~\cite{Gal'tsov:2008km}).

The semi-classical dynamics of the homogeneous $SU(2)$ condensate with small (but non-zeroth) YM wave modes
has been thoroughly studied in Minkowski spacetime in~\cite{Prokhorov:2013xba}. The results reveal the characteristic
decay of the YM condensate, such that its energy gets effectively transferred from the condensate to the YM wave modes
heating up the ultra-relativistic YM plasma. In principle, this effect can be relevant, for example, for a better understanding
of the particle production mechanisms in the cosmological plasma and could, potentially, force the inflationary stage
driven by the YM condensate to terminate. The additional spatially-inhomogeneous quantum-wave contributions to
the QCD vacuum can be associated with spatial averages of the higher dimensional operators, such as
$\langle (A_\mu^a)^2 \rangle$, in the semi-classical treatment. Such averages describe the contribution
of the hadron modes to the ground state, particularly relevant after the QCD phase transition, and, thus,
should be incorporated into the analysis. The latter aspects, however, go beyond the considered classical limit
and should be explored in the effective YM theory at the quantum level.

The major role of quantum effects on the dynamics of the Born--Infeld (BI) field condensate in cosmology
similar to those in QCD has been discussed in~\cite{Elizalde:2003ku}. Namely, it was shown
that the quantum corrections leading to a non-zeroth trace of the energy-momentum tensor
of the BI field affect the long-range behavior of the BI condensate introducing time-dependent
corrections to the energy and pressure of the BI field, thereby altering its equation of state away
from that of a classical radiation fluid. This motivates a deeper study of quantum effects on the dynamics
of the ground state in YM theories.

\section{Yang--Mills Effective Action}

As was demonstrated in~\cite{Savvidy}, the classical YM equations of
motion following from the classical action (\ref{L}) are form non-invariant with
respect to infinitesimally small quantum fluctuations breaking the conformal invariance
of the gauge theory. The latter effect is known as a conformal anomaly, which has notable
consequences in cosmology. Indeed, any infinitesimal external field affects the classical YM vacuum by
modifying the initial operator (non-linear) YM equations, since there is not
any physical threshold for the vacuum polarization of a massless quantum
YM field by its classical component. That results in the well-known fact
that the solutions of the YM equations are unstable w.r.t radiative corrections
and cannot be used in physical applications. In practice, we work with
the so-called Savvidy vacuum fluctuations and look for their spatially-homogeneous modes.
In order to construct the realistic EYM equations describing YM condensate
dynamics in a non-stationary background of the expanding Universe, one must consistently
incorporate, at least, the lowest-order corrections from the vacuum polarization
in the effective YM Lagrangian.

Consider the generic approach, which leads to the YM energy-momentum tensor incorporating
conformal anomalies. According to this approach, when applying the variational procedure in the derivation
of the YM equations, the gauge coupling $g_{\rm YM}$ is treated as an operator depending on
the quantum fields' operators. The gauge field operator $\mathcal{A}_\mu^a$ is then considered
as a variational variable, which together with the corresponding stress tensor operator is related
to those in the standard normalization as in Equation~(\ref{AF}). The effective action and Lagrangian
operators of the quantum gauge theory are given in terms of the gauge-invariant operator of the least
dimension $J$, which plays the role of an order parameter for the YM condensate (a combination
of magnetic $\vec B$ and electric $\vec E$ field contributions) by \cite{Savvidy,Pagels,Adler:1981as}:
\begin{equation}
S_{\rm eff}[\mathcal{A}] = \int \mathcal{L}_{\rm eff} d^4x \,, \qquad \mathcal{L}_{\rm
eff}=-\frac{J}{4g_{\rm YM}^2(J)}\,, \qquad J=\mathcal{F}^2\equiv \mathcal{F}^a_{\mu\nu}\mathcal{F}_a^{\mu\nu}=2(B^2-E^2)\,,
\label{Lrg}
\end{equation}
respectively, whose variation w.r.t $\mathcal{A}_\mu^a$ leads to the operator
energy-momentum tensor of the gauge~theory:
\begin{equation}
\displaystyle
\hat{T}_{\mu}^{\nu,{\rm YM}}=\frac{1}{g_{\rm YM}^2}\left(-\mathcal{F}^a_{\mu\sigma}\mathcal{F}_a^{\nu\sigma}
+\frac14\delta_\mu^\nu \mathcal{F}^a_{\rho\sigma}\mathcal{F}_a^{\rho\sigma}+\frac{\beta(g_{\rm YM}^2)}{2}
\mathcal{F}^a_{\mu\sigma}\mathcal{F}_a^{\nu\sigma}\right) \,,
\label{S-tem}
\end{equation}
providing the usual form of the trace anomaly relation:
\begin{eqnarray}
\displaystyle
\hat{T}_{\mu}^{\mu,{\rm YM}}=\frac{\beta(g_{\rm YM}^2)}{2g_{\rm YM}^2}\,J \,, \qquad g_{\rm YM}^2=g_{\rm YM}^2(J) \,.
\end{eqnarray}
The gauge coupling dependence on $J$ is determined by the RG
 evolution equation in the following operator form:
\begin{equation}
\displaystyle 2J\frac{dg_{\rm YM}^2}{dJ}=g_{\rm YM}^2\,\beta(g_{\rm YM}^2) \,,
\label{rg}
\end{equation}
where $\beta=\beta(g_{\rm YM}^2)$ is the standard $\beta$-function. Note, the RG
Equation (\ref{rg}) is symmetric with respect to $J \leftrightarrow -J$; thus, its solution is determined
by the absolute value of the YM invariant operator, \textit{i.e.}:
\begin{eqnarray}
g_{\rm YM}^2=g_{\rm YM}^2(|J|) \,.
\end{eqnarray}
which has important consequences on stability of the ground-state YM solutions
in Minkowski spacetime. On should therefore consider the effective action (\ref{Lrg})
as a classical model \cite{Pagels}, which possesses well-known properties of the
full quantum theory, such as (i) local gauge invariance (ii) RG evolution and asymptotic
freedom, (iii) correct quantum vacuum configurations and (iv) trace anomaly.
These provide a sufficient motivation and physics interest in cosmological aspects
of considering the model effective.

The Perturbation Theory 
 can be applied to the effective action in the limit of
large mean fields, {\it i.e.}, $|J|\to \infty$, away from the classical ground state.
To the one-loop approximation, the solution of the RG Equation (\ref{rg})~reads:
\vspace{-6
pt}
\begin{equation}
\displaystyle \beta(g_{\rm YM}^2)=-\frac{bg_{\rm YM}^2}{16\pi^2}\,, \qquad
g_{\rm YM}^2=\frac{32\pi^2}{\displaystyle b\ln(|J|/\lambda^4)}\,, \label{be}
\end{equation}
where $\lambda$ is the scale parameter and $b$ is the one-loop $\beta$-function
coefficient (e.g., in pure $SU(3)$ gauge theory $b=11$).
Substituting this solution into the effective Lagrangian (\ref{Lrg})
and performing a straightforward covariant generalization for a curved background with
metric $g_{\mu\nu}$ ($g\equiv {\rm det}(g_{\mu\nu})$), we obtain finally \cite{Savvidy:1977as}:
\begin{equation}
\mathcal{L}^{\rm 1-loop}_{\rm eff}=-\frac{b\,J}{128\pi^2}\,
\ln\Big(\frac{|J|}{(\xi\lambda)^4}\Big) \,, \qquad
J=\frac{\mathcal{F}^a_{\mu\nu}\mathcal{F}_a^{\mu\nu}}{\sqrt{-g}} \,,
\label{Lrg-1}
\end{equation}
where free parameter $\xi$ reflects an arbitrariness in the multiplicative normalization of the invariant $J$.
Both parameters $\xi$ and $\lambda$ are not fixed by the theory, but can be determined from phenomenology
in realistic gauge theories, such as QCD, where $\lambda \equiv \Lambda_{\rm QCD} \simeq 280$ MeV.
The effective action (\ref{Lrg-1}) or rather its non-perturbative generalization (see below) can then be considered
as a classical model incorporating important features of the full quantum model and providing the correct description
of the quantum vacuum \cite{Pagels}. In what follows, we apply classical methods to study its physically-relevant~configurations.

In asymptotically free gauge theories like QCD, the quantum vacuum configurations are controlled by the strong coupling regime.
Performing an analysis in Euclidean spacetime, in~\cite{Pagels}, it was shown that the vacuum value of the gauge invariant
$\langle J \rangle$ in a strongly-coupled quantum gauge theory does not vanish as it does in the classical gauge theory, and
the corresponding functional integral is not dominated by the minima of the classical action (\ref{FuncI}).
Moreover, it was shown that there are no instanton solutions to the effective action (\ref{Lrg}), such that the ground state of
the quantum YM theory does not contain the classical instanton configurations. Instead, the quantum vacuum within the effective
action (\ref{Lrg}) approach can be understood as a state with ferromagnetic properties, which undergoes the spontaneous
magnetization, providing a consistent description of the non-perturbative QCD vacuum. Thus, the quantum effects drastically
affect the properties of the ground state in YM theories, which may have profound consequences in cosmology.

\section{Cosmological Evolution of the Yang--Mills Ground State}

The condensation of YM fields has been first considered as a source of the cosmic inflation in~\cite{Zhang:1994pm}
in the framework of the RG-improved effective action (\ref{Lrg}). For $|J|\gg (\xi \lambda)^4$, one reaches the asymptotically-free regime
where the one-loop approximated effective model (\ref{Lrg-1}) can be considered to be reliable. In this limit, the EYM equations exhibit
a radiation-dominated solution $a(t)\propto t^{1/2}$, $p_{\rm YM}=\epsilon_{\rm YM}/3$, which is characteristic for the classical YM behavior considered above.
The effective Lagrangian (\ref{Lrg-1}) attains the minimum at $|J_0| = {\rm const}$ corresponding to the de-Sitter solution $a(t)\propto \exp(Ht)$,
with the vacuum equation of state $p_{\rm YM}=-\epsilon_{\rm YM}$, which can be applied in the context of early-time acceleration (provided that
a new strongly-coupled high-scale dynamics does exist in Nature). The de-Sitter (vacuum) solution appears to be stable with respect to small perturbations
of the scale factor and the YM condensate for $J_0<0$, which are exponentially decaying with time. 
These ideas have been further applied for the current cosmic acceleration, as a source of DE, in the one-loop effective model (\ref{Lrg-1}) 
in~\cite{Zhao:2006mk}, while the main aspects of first-order cosmological perturbation theory with the YM condensate have been 
discussed in~\cite{Zhao:2005ap}. The corresponding one-loop solution has been further analyzed in~\cite{Pasechnik:2013sga} 
and employed for the possible compensation of the non-perturbative QCD vacuum contribution (\ref{Lambda-QCD}). 
Let us discuss the key features of this model in more detail.

\subsection{Exact Solutions of the One-Loop Effective Model}

Assuming that the perturbative vacua are already eliminated at the typical time scale of the QCD transition,
the EYM equations of motion for the effective QCD theory (\ref{Lrg-1}) without taking into
account the gluon field interactions with usual matter can be written as follows:
\begin{small}
\begin{eqnarray}
&&\frac{1}{\kappa}\left(R_\mu^\nu-\frac12\delta_\mu^\nu R\right)=
\frac{b}{32\pi^2}\frac{1}{\sqrt{-g}}
\biggl[\biggl(-\mathcal{F}_{\mu\lambda}^a\mathcal{F}^{\nu\lambda}_a
+\,\frac14\delta_\mu^\nu
\mathcal{F}_{\sigma\lambda}^a\mathcal{F}^{\sigma\lambda}_a\biggr)
\ln\frac{e|\mathcal{F}_{\alpha\beta}^a\mathcal{F}^{\alpha\beta}_a|}{\sqrt{-g}\,
(\xi\lambda)^4}-\frac14 \delta_\mu^\nu \,
\mathcal{F}_{\sigma\lambda}^a\mathcal{F}^{\sigma\lambda}_a\biggr]\,,\label{maineq}
\\
&& \left(\frac{\delta^{ab}}{\sqrt{-g}}\partial_\nu\sqrt{-g}-f^{abc}\mathcal{A}_\nu^c\right)
\left(\frac{\mathcal{F}_b^{\mu\nu}}{\sqrt{-g}}\,\ln\frac{e|\mathcal{F}_{\alpha\beta}^a
\mathcal{F}^{\alpha\beta}_a|}{\sqrt{-g}\,(\xi\lambda)^4}\right)=0\,,\nonumber
\end{eqnarray}
\end{small}
where $e$ is the base of the natural logarithm. In the FLRW Universe, the equations (\ref{maineq}) for
the homogeneous $SU(3)$ gluon condensate $U=U(\eta)$ and the expansion law $a=a(\eta)$ straightforwardly
transform to:
\begin{eqnarray}
&& \frac{6}{\kappa} \frac{a''}{a^3} = \frac{3b}{16\pi^2 a^4}\Big[(U')^2-\frac{1}{4}U^4\Big]\,, \nonumber
\\
&& \frac{\partial}{\partial \eta}\Big(U'\,\ln\frac{6e\big|(U')^2-\frac{1}{4}U^4\big|}{a^4(\xi \lambda)^4}\Big) +
\frac{1}{2}U^3\,\ln\frac{6e\big|(U')^2-\frac{1}{4}U^4\big|}{a^4(\xi \lambda)^4}=0 \,. \label{eqU}
\end{eqnarray}
The first integral of this system reads:
\begin{eqnarray}
&& \frac{3}{\kappa}\frac{(a')^2}{a^4}=\frac{3b}{64\pi^2 a^4}\,\Big(\Big[(U')^2+\frac{1}{4}U^4\Big]\,
\ln\frac{6e\big|(U')^2-\frac{1}{4}U^4\big|}{a^4(\xi \lambda)^4} + (U')^2 - \frac{1}{4}U^4\Big)\,, \label{eqUint}
\end{eqnarray}
which yields two exact partial solutions satisfying:
\begin{eqnarray}
Q(\eta)\equiv \frac{6e\big[(U')^2-\frac{1}{4}U^4\big]}{a^4(\xi \lambda)^4}=\pm 1\,. \label{Uexact}
\end{eqnarray}
providing contributions to the CC:
\begin{eqnarray}
\Lambda^{\rm QCD}_\pm = \pm \frac{3b}{64\pi^2}\,\frac{(\xi \Lambda_{\rm QCD})^4}{6e} \,. \label{YM-dens-an}
\end{eqnarray}
in the QCD case corresponding to chromoelectric ($E^2>B^2$) and chromomagnetic ($E^2<B^2$) condensates, respectively
\cite{Zhang:1994pm}. Indeed, these are special solutions for which the quantum and classical traceless parts of the YM
energy-momentum tensor are mutually canceled.

In fact, the exact solutions (\ref{Uexact}) are the special attractor (or tracker) solutions for the general solution of the EYM system (\ref{eqUint})
(see, e.g.,~\cite{Zhao:2006mk}). Indeed, a close consideration demonstrates that irrespective of the initial value of the function $\tilde{Q}(t)\equiv Q(\eta(t))$
at some initial moment in physical time $t=t_0$, the YM system approaches the state given by one of the two exact solutions (\ref{Uexact}), {\it i.e.}:
\begin{eqnarray}
&& \tilde{Q}(t_0) > 0\,, \qquad \tilde{Q}(t)|_{t \to \infty} \to +1 \,, \label{sol-I}
\\
&& \tilde{Q}(t_0) < 0\,, \qquad \tilde{Q}(t)|_{t \to \infty} \to - 1 \,. \label{sol-II}
\end{eqnarray}
It is tempting to identify the negative-energy solution $\Lambda^{\rm QCD}_{-}$ with the phenomenologically known
value of the negative-valued quantum-topological contribution (\ref{Lambda-QCD}):
\begin{eqnarray}
\Lambda^{\rm QCD}_{-} \equiv \Lambda^{\rm QCD}_{\rm vac} \qquad \to \qquad \xi \simeq 4 \,, \label{YM-top}
\end{eqnarray}
which fixes the arbitrary normalization factor $\xi$ of the invariant $J$. On the other hand, notice that spatially-homogeneous and
topological subsystems of the QCD vacuum have an entirely different nature and cannot be treated on the same footing.
For both the general and partial solutions, the amplitude of the gluon condensate $U(t)$ oscillates with quasiperiodic singularities
in physical time, whereas its energy density and pressure remain continuous. This is in variance to the continuous classical
YM solution (\ref{classol}).

\subsection{QCD Vacuum Compensation}

The asymptotic regimes (\ref{sol-I}) and (\ref{sol-II}) are reached after a number of oscillations of the gluon condensate
density and pressure, and such a transition is accompanied by a decelerating expansion of the Universe (filled by
the gluon condensate only). The characteristic (relaxation) time scale for such a transition reads:
\begin{eqnarray}
t_{rel} \simeq \frac{1}{\sqrt{\kappa \epsilon^{\rm QCD}_0}} \,, \qquad \epsilon^{\rm QCD}_0 \equiv \epsilon^{\rm QCD}(t=t_0)
\gg \Lambda_{\rm cosm} \,,
\end{eqnarray}
where the observed $\Lambda_{\rm cosm}$ is given by Equation~(\ref{Lcosm}), and $\epsilon^{\rm QCD}_0>0$ is the initial energy of the YM condensate.
Hypothetically, both subsystems with positively- (electric) and negatively- (magnetic) definite energy could have been generated at the QCD phase
transition epoch and then asymptotically (at $t\gg t_{rel}$) approach their constant values $\Lambda^{\rm QCD}_{+}$ and
$\Lambda^{\rm QCD}_{-}=-\Lambda^{\rm QCD}_{+}$, respectively. If both types of the YM condensate really co-exist
in the QCD vacuum, they would provide an automatic compensation of the net QCD vacuum energy in the IR limit without any fine-tuning.
Such a compensation can be viewed as a manifestation of the QCD confinement, since there are practically no non-zeroth gluon
fields propagating at the length scales larger than the typical hadron scale $\sim1$ fm, and they certainly disappear at macroscopically
large cosmological scales typical for the modern~Universe.

A similar idea about the heterogenic structure of the QCD vacuum containing different co-existing vacuum subsystems
at the $\Lambda_{\rm QCD}$ scale, which may eliminate each other under certain conditions, was discussed in~\cite{Pasechnik:2013poa}.
In particular, besides the collective quark-gluon excitations of a topological nature, the QCD ground state should contain
long-range quantum-wave excitations of lightest bound states (hadrons), such that both subsystems overlap and potentially
compensate each other in the IR limit of the theory. Indeed, the contribution from the collective quantum-wave fluctuations
to the net QCD vacuum energy as estimated phenomenologically is \cite{Shifman-1,Shifman-2}:
\begin{eqnarray}
\epsilon_{\rm vac(h)}=\frac{1}{32\pi^2}\left(2\sum_B
(2J_B+1)m_B^4\ln\frac{\mu}{m_B} - \sum_M
(2J_M+1)m_M^4\ln\frac{\mu}{m_M}\right) \,,
\end{eqnarray}
where $\mu$ is a upper momentum cut-off parameter and $J_B$ and $J_M$ ($m_B$ and $m_M$) are the spins (masses) of
lightest baryon and meson degrees of freedom, respectively. Taking into account only metastable hadronic degrees of
freedom---the baryon octet $B=\{N,\,\Lambda,\,\Sigma,\,\Xi\}$ and the pseudoscalar nonet $M=\{\pi,\,K,\,\eta,\,\eta'\}$---one obtains the desired result, namely,
\begin{eqnarray}
\Lambda_{\rm vac}^{\rm QCD} + \epsilon_{\rm vac(h)} = 0\,, \qquad \mu\simeq 1.2 \;\text{GeV} \,,
\end{eqnarray}
{\it i.e.}, topological and wave fluctuations contribute to the QCD vacuum energy with opposite signs and may, in principle, compensate
each other, which, however, requires a significant fine-tuning. Therefore, the exact compensation of the QCD vacuum may be a universal
phenomenon, which is present in various treatments of the QCD vacuum. The picture with the apparent asymptotic cancellation of
chromoelectric and chromomagnetic vacuum components discussed above, if realized in Nature, may have profound advantages
compared to the phenomenological models due to an attractor nature of the corresponding solutions and the absence of any fine tuning.

In fact, exact compensation of energies of the vacuum components does not mean that the sum of their quantum fluctuations,
which may have very different spacetime dynamics, is identically equal to zero. The complete (unordered) two-point function
in the QCD vacuum can then be represented as a superposition of two parts:\vspace{-4pt}
\begin{eqnarray}
\langle0|\frac{\alpha_s}{\pi}F^a_{ik}(x)F_a^{ik}(x')|0\rangle\equiv
\langle0|\frac{\alpha_s}{\pi}F^a_{ik}(x)F_a^{ik}(x')|0\rangle_{(+)} +
\langle0|\frac{\alpha_s}{\pi}F^a_{ik}(x)F_a^{ik}(x')|0\rangle_{(-)} \,,
\label{xx'}
\end{eqnarray}
where $+$ and $-$ refer to the contributions (\ref{sol-I}) and (\ref{sol-II}), respectively,
which are represented:
\begin{eqnarray*}
&& \langle0|\frac{\alpha_s}{\pi}F^a_{ik}(x)F_a^{ik}(x')|0\rangle_{(+)} =
\langle0|:\frac{\alpha_s}{\pi}F^a_{ik}(0)F_a^{ik}(0):|0\rangle\, D_{(+)}(x-x') \,, \\
&& \langle0|\frac{\alpha_s}{\pi}F^a_{ik}(x)F_a^{ik}(x')|0\rangle_{(-)} =
\langle0|:\frac{\alpha_s}{\pi}F^a_{ik}(0)F_a^{ik}(0):|0\rangle\, D_{(-)}(x-x') \,,
\end{eqnarray*}
in terms of the experimentally-measured local condensate \cite{Shifman-1,Shifman-2},
\begin{eqnarray}
\langle0|:\frac{\alpha_s}{\pi}F^a_{ik}(0)F_a^{ik}(0):|0\rangle=(1.7\pm 0.4)\times 10^{-2}\,\rm{GeV}^4\,, \label{top1}
\end{eqnarray}
and correlation functions $D_{(\pm)}(x)$ satisfying $D_{(\pm)}(0)=1$.
The latter should be constrained by non-perturbative methods, e.g., lattice
QCD or by means of effective field theory methods. Finally,
the complete non-local condensate can be written as:
\begin{eqnarray} \nonumber
&& \langle0|\frac{\alpha_s}{\pi}F^a_{ik}(x)F_a^{ik}(x')|0\rangle=
\langle0|:\frac{\alpha_s}{\pi}F^a_{ik}(0)F_a^{ik}(0):|0\rangle
D(x-x')\,, \\
&& D(x-x')\equiv D_{(+)}(x-x')-D_{(-)}(x-x')\,, \qquad D(0)=0 \,.
\label{dcomp}
\end{eqnarray}
Below, we make use of this important relation in the calculation of
the gravitational correction to the ground state energy in QCD.

\subsection{Asymptotic Behavior of the QCD Vacuum Energy}

As long as such an asymptotic regime is achieved, the macroscopic evolution of the Universe reduces to the standard (slow)
Friedmann evolution driven only by a small observed residue $\Lambda_{\rm cosm}$, as well as by all other (non-DE) forms
of matter with density $\epsilon_{\rm mat}$, {\it i.e.}:
\begin{eqnarray}
\frac{3}{\kappa}\frac{(a')^2}{a^4}=\epsilon_{\rm mat}+\Lambda_{\rm cosm}\,,
\end{eqnarray}
and the rapid microscopic evolution of the gluon condensate $U=U(\eta)$ at the characteristic QCD time scale:
\begin{eqnarray}
(U')^2-\frac{1}{4}U^4=a^4 \frac{(\xi\Lambda_{\rm QCD})^4}{6e} \,,
\end{eqnarray}
practically decoupled from the slow macroscopic Universe expansion. In the limit $t\gg t_{rel}$, the positive- and
negative-valued components of the gluon condensate evolve as:
\begin{eqnarray}
\int_{\widetilde{U}_0}^{\widetilde{U}}\frac{dx}{\sqrt{\frac{1}{4}x^4 \pm 1}}=\widetilde{\eta}\,, \qquad
\widetilde{U}=U\frac{(6e)^{1/4}}{\xi\Lambda_{\rm QCD}}\,,\qquad
\widetilde{\eta}=\eta\frac{\xi \Lambda_{\rm QCD}}{(6e)^{1/4}}\,, \qquad \xi \simeq 4\,, \label{Uexact2}
\end{eqnarray}
respectively.
As was shown in~\cite{Pasechnik:2013sga}, the cosmological evolution of the gluon field in its ground
state can be interpreted as a regular sequence of quantum tunneling transitions through the ``time barriers''
represented by the regular singularities in the quantum vacuum solution of the effective model (\ref{Lrg-1}).
Such a behavior is analogical to that of the Dirac monopole where a singularity in the four-potential emerges along the Dirac string
while the magnetic flux is finite. The position of the Dirac string depends on the gauge choice. Similarly, the positions of
the singularities in the $U(t)$ amplitude depend on the gauge choice and correspond to the regular minima of the oscillating
trace of the energy-momentum tensor. A key difference is that in the case of magnetic monopole, the singularity appears along
spatial directions, while in the case of the classical YM condensate, its potential is singular at discrete points in time.

At the intermediate time scale corresponding to the modern Universe,
\begin{eqnarray}
\label{today}
t_{rel}\ll t \sim \frac{1}{\sqrt{\kappa \Lambda_{\rm cosm} }}
\end{eqnarray}
the exact compensation between the two components of the gluon condensate may not be reached yet. Therefore, let us assume
that the DE is not a cosmological constant and that it is entirely given by a time-dependent uncompensated (positive) contribution
from the gluon condensate, such that:
\begin{eqnarray}
\label{presc-1}
\delta\epsilon(t)\equiv \epsilon^{\rm QCD}_{(+)}(t) + \epsilon^{\rm QCD}_{(-)}(t) >0 \,, \qquad \delta\epsilon(t) \ll \epsilon^{\rm QCD}_{(\pm)}(t) \,,
\end{eqnarray}
such that the exact compensation condition $\delta\epsilon(t)\to 0$ is satisfied for $t\to \infty$. Besides, assume for simplicity
that in the modern Universe, the negative-valued (magnetic) component is much closer to its asymptotic state (\ref{sol-II}):
\begin{eqnarray}
\label{presc-1-1}
\Delta \epsilon_{(+)} \gg \Delta \epsilon_{(-)}\,, \qquad \Delta \epsilon_{(\pm)}\equiv |\epsilon^{\rm QCD}_{(\pm)}(t) - \Lambda^{\rm QCD}_{\pm}| \,,
\end{eqnarray}
then the positive (electric) one is such that the sum of the two contributions is positive today, as required by observations, and in the past.
In this case, the DE component is almost entirely dominated by the electric component (minus its asymptotic
value $\Lambda^{\rm QCD}_{+}$). Then, a straightforward asymptotic analysis of the general solution of Equation~(\ref{eqU}) uncovers
that at the time scale corresponding to the modern Universe (\ref{today}), the cosmological expansion, as well as the net gluon condensate
density and pressure evolve with time as:
\begin{eqnarray}
\label{oneloop-res}
a(t)\simeq a^* \Big( \frac{3\kappa \epsilon^{\rm QCD}_0}{4} \Big)^{1/3}\, t^{2/3}\,, \qquad
\delta\epsilon(t) \simeq \frac{4}{3\kappa t^2}\,, \qquad p^{\rm QCD}(t) = -\frac13\delta\epsilon(t)\,\chi(t)\,,
\end{eqnarray}
where $\chi(t)$ is an auxiliary function satisfying a non-linear differential equation:
\begin{eqnarray}
\dot{\chi}^4=\frac83\,(\xi \Lambda_{\rm QCD})^4\,(1-\chi^2)^3 \,.
\end{eqnarray}
This function rapidly oscillates about zero with period $\tau \sim \Lambda_{\rm QCD}^{-1}$ and with unit amplitude.
The equation of state for the gluon condensate averaged over many microscopic oscillations of pressure in
the modern Universe $t=t_U$:
\begin{eqnarray}
\label{eos}
\langle p^{\rm QCD}(t_U) \rangle=0\,, \qquad w=\frac{\langle p^{\rm QCD}(t_U) \rangle}{\delta\epsilon(t_U)} = 0
\end{eqnarray}
corresponding to the case of dust-type matter and, therefore, does not provide a candidate for a traditional vacuum-like CC
in the standard cosmological model, in variance to~\cite{Zhao:2009wy}. Note, the energy-density $\delta\epsilon(t)$ of such
a matter does not depend on initial conditions, and today it amounts~to:
\begin{eqnarray}
\label{DEnow}
\delta\epsilon(t_U)\simeq \frac{4}{3\kappa t_U^2} \simeq 1.8 \times 10^{-47}\, {\rm GeV}^4 \,,
\end{eqnarray}
which is not far from the measured value (\ref{Lcosm}), although it does not immediately fit the observations, and it is not entirely
clear if this result poses a big problem or yet another DE candidate for modern cosmology. Note, in~\cite{Shaw:2010pq,Barrow:2010xt},
in a generic approach based on the extended Einstein--Hilbert variational principle, the observed $\Lambda_{\rm cosm}$ is promoted to
a ``field''. Such a procedure leads to a consistent estimate $\Lambda_{\rm cosm}\sim 1/t^2$ provided that the modified variational
approach is indistinguishable from GR with a CC having a value $1/t^2$ at the moment of observations $t$ similarly to Equation~(\ref{DEnow}),
although the equation of state is argued to be of a CC/vacuum type.

Of course, the presence of the gluon condensate in the early Universe is unavoidable, and its long-range implications have to be studied
with special care. In particular, the dynamics of quantum waves yet remain to be incorporated, and an additional effect of parametric resonance
causing the condensate to decay into particles \cite{Prokhorov:2013xba} should be studied in the effective action approach. Such a possibility
for the condensate decay would, in principle, reduce or even eliminate the naive estimate (\ref{DEnow}), as well as it could affect the equation of
state (\ref{eos}). Besides, the simple toy-model above does not account for the higher-order perturbative and non-perturbative QCD effects,
as well as additional possible contributions to the DE/CC of a different nature. These big open questions offer a large space for further explorations.

\subsection{Towards Non-Perturbative QCD}

An improved perturbative analysis of the DE for the effective YM action at two- and three-loop levels has been performed
in~\cite{Xia:2007eu} and \cite{Wang:2008fx}, respectively (see also~\cite{Zhao:2009wy}). Remarkably,
the perturbative quantum corrections do not spoil the attractor nature of the DE solutions, and the theory retains
its asymptotic properties and stability. However, for the exact solutions of the EYM systems (\ref{sol-I}) and (\ref{sol-II}),
the perturbative expansion for the $\beta$-function apparently breaks down, which naturally raises an important question about
the validity of straightforward extrapolations of the perturbative effective theory results into the deeply non-perturbative domain
(this issue has also been discussed in~\cite{Pagels}). In order to address this issue, it is critical to extend the above analysis
beyond the perturbative fixed-order approximation and to check if the observed asymptotic character of the theory with analogical
CC solutions is preserved there or not.

This analysis has been performed recently in the context of the effective non-perturbative YM condensate DE in~\cite{Dona:2015xia}.
Using non-perturbative techniques of the functional renormalization group~\cite{Eichhorn:2010zc}, it was shown that the effective Lagrangian
(\ref{Lrg}) has a minimum in the order parameter $J$, which provides the existence of the attractor CC solutions analogical to
Equations~(\ref{sol-I}) and (\ref{sol-II}) and can reproduce the DE behavior of the expanding Universe as small redshifts (with and without
interactions with external matter fields). It turns out that the generic non-perturbative effective QCD theory inherits properties similar
to the fixed-order perturbative models. Namely, in the asymptotically-free regime in the early Universe, the non-perturbative gluon condensate
behaves as a radiation medium with $w=1/3$ consistent with the classical YM evolution, while at later time scales (depending on initial conditions),
the condensate falls into the critical state with $w=-1$ being an attractor for a general solution, as was seen in the one-loop toy-model discussed above.

For illustration, consider the energy momentum tensor of the gluon condensate without imposing any perturbative constraints
on the $\beta$-function (see also~\cite{Pagels}):
\begin{eqnarray}\nonumber
T^{\nu,{\rm YM}}_{\mu}= - \Big[1-\frac12\beta(g^2_{\rm YM})\Big]\frac{\mathcal{F}^a_{\mu\lambda}
\mathcal{F}_a^{\nu\lambda}}{g_{\rm YM}^2\,\sqrt{-g}} + \delta_{\mu}^{\nu}\frac{J}{4 g^2_{\rm YM}}\,,
\label{T}
\end{eqnarray}
where $g^2_{\rm YM}=g^2_{\rm YM}(|J|)$ and $J$ is given in covariant form in Equation~(\ref{Lrg-1}).
The all-order YM equation of motion then reads:
\begin{eqnarray}
\left(\frac{\delta^{ab}}{\sqrt{-g}}\partial_\nu\sqrt{-g}-f^{abc}\mathcal{A}_\nu^c\right)
\left[\frac{\mathcal{F}_b^{\mu\nu}}{g_{\rm YM}^2\,\sqrt{-g}}\,
\Big(1-\frac12\beta\big(g^2_{\rm YM}\big)\Big)\right]=0\,.
\label{YMeq}
\end{eqnarray}
This equation has a simple manifestly non-perturbative solution:
\begin{eqnarray}
\label{SS}
\beta\big(g^2_{\rm YM}(|J|)\big)=2 \,,
\end{eqnarray}
which should be realized asymptotically in the IR limit, in the case of QCD, soon after the Universe temperature drops below the QCD
transition temperature $T_{\rm QCD}\sim 0.1$ GeV.

In practice, we do not know the form of the non-perturbative $\beta$-function. The solution (\ref{SS}) either fixes the invariant
$J$ to its constant initial value $J_0=J(t=t_0)$ or, alternatively, indicates that the non-perturbative $\beta$-function is constant
and does not depend on the invariant $J$ and, hence, on $g^2_{\rm YM}$ in the strong coupling regime. In both cases, the solution (\ref{SS})
is a non-perturbative analog of the exact solution (\ref{Uexact}), since it eliminates the quantum and classical traceless
parts of the YM energy-momentum tensor (\ref{T}).

As a particularly simple example, consider the case of running invariant $J$ and constant non-perturbative $\beta=2$. Physically, the latter
possibility would correspond to the saturated form of the function $\beta=\beta(g_{\rm YM}^2)$, which approaches two
at large coupling $g_{\rm YM}>1$ where the attractor solution is concerned. Then, the energy-density of the YM condensate and
the non-perturbative RG equation read:
\begin{eqnarray}
\epsilon^{\rm YM}= \frac{J}{4 g^2_{\rm YM}(|J|)}\,, \qquad
\frac{d\ln g_{\rm YM}^2}{d\ln(|J|/\lambda^4)}= 1 \,,
\label{T00-GS}
\end{eqnarray}
implying that:
\begin{eqnarray}
g_{\rm YM}^2(|J|)=g^2_0\Big|\frac{J}{J_0}\Big|>0\,, \qquad g^2_0\equiv g_{\rm YM}^2(|J_0|) \,, \qquad J_0=J(t=0)\,.
\label{J0}
\end{eqnarray}
Therefore, the YM condensate again emerges as a CC in the IR limit of the effective theory:
\begin{eqnarray}
T^{\nu,{\rm YM}}_{\mu}=\Lambda_{\rm YM}\delta_{\mu}^{\nu}\,, \qquad \Lambda_{\rm YM}\equiv
\pm \frac{|J_0|}{4 g^2_0}\,,
\label{Tsol}
\end{eqnarray}
whose sign is determined by the sign of the invariant $J$. Exactly the same solutions emerge in the case of a varying $\beta$-function
and a fixed $J_0$ value for which the effective Lagrangian (\ref{Lrg}) is minimal. Apparently, these solutions are qualitatively
the same as the one-loop attractor solutions (\ref{sol-I}) and (\ref{sol-II}), and the precise form of the non-perturbative
$\beta$-function was not relevant to justify that. Remarkably, a precise value of the $J_0$ and $g^2_0$ parameters is not relevant
as long as the asymptotic value $\Lambda_{\rm YM}$ is subtracted, since the resulting energy density of the gluon condensate
contribution to the modern DE (\ref{DEnow}) does not depend on initial conditions. One should further investigate the equation of
state and time dependence of the resulting DE density in the non-perturbative case in order to make a final conclusion about the stability
of the results of the one-loop effective model (\ref{oneloop-res}).

\section{The Role of Gravity: Zeldovich--Sakharov Scenario}

Within the traditional QFT-based approaches, there are promising attempts to address the smallness of
the observable $\Lambda$-term density value by interpreting it as a quantum gravity correction to the ground state energy,
{\it i.e.}, by treating the positive DE density as a small, but non-vanishing effect of gravitating non-perturbative
vacuum fluctuations in the expanding Universe.

Long ago, Zeldovich had pointed out in~\cite{Zeldovich:1967gd} that the $\Lambda$-term density gets contributions from graviton-exchange
interactions between virtual elementary particles in the physical vacuum providing $\Lambda_{\rm cosm}\sim G m^6$, where $m$ is some characteristic
mass of light particles. Sakharov has also noticed in~\cite{Sakharov-1,Sakharov-2} that extra terms describing an effect of graviton exchanges between
identical particles (e.g., bosons in the ground state) should appear in the right-hand side of Einstein equations averaged over their quantum
ensemble. Even before the modern CC value has become known, in~\cite{Kardashev}, the Zeldovich relation has been represented through
the basic fundamental constants, the minimal (typical hadron scale) and maximal (Planck mass) fundamental scales as follows:
\vspace{-4pt}
\begin{eqnarray}
 \Lambda_{\rm cosm}=\frac{m_\pi^6}{(2\pi)^4M^2_{\rm Pl}}\simeq
 3.0\times 10^{-47}\; \text{GeV}^4\,, \label{Lambda-mpi}
\end{eqnarray}
where $m_\pi \simeq 138\;\text{MeV}$ is the pion mass. It is worth noticing that the representation (\ref{Lambda-mpi})
turns out to be numerically close to the observed CC value (\ref{Lcosm}), a rather interesting coincidence, which has triggered further studies
in the literature (for a more detailed review on this topic, see~\cite{Sola:2013gha}).

Along these lines, a recent approach of~\cite{Klinkhamer:2009nn} is based on a generic ``$q$-theory'' operating with a conserved microscopic
$q$ value, whose statics and dynamics are studied at the macroscopic scales. Such a quantity can, in principle, be identified with the gluon
condensate in QCD, which naturally gravitates, resulting in a nonzero DE value in the non-equilibrium state of the expanding Universe
estimated as $\Lambda_{\rm cosm} \propto \Lambda_{\rm QCD}^6/M_{\rm PL}^2$. A similar estimate was explained in~\cite{Thomas:2009uh} within the effective gravity coupled to the QCD sector via the trace anomaly.

Another approach of~\cite{Urban:2009yg} considers the dynamics of the ghost fields in the low energy (chiral) QCD and its dynamical
effects in a spacetime with a non-trivial topological structure. In particular, the Veneziano ghost \cite{Veneziano:1979ec}, which is unphysical
in the standard Minkowski QFT, results in a non-vanishing physical effect in the expanding Universe parametrized by a deviation from
the Minkowski vacuum, in a similar way to the Casimir energy:
\begin{eqnarray} \label{renL}
\Delta \epsilon_{\rm vac} \equiv \epsilon_{\rm FLRW} - \epsilon_{\rm Mink} \,,
\end{eqnarray}
which appears to be naturally small and depends on the properties of the external gravitational background only \cite{Schutzhold:2002pr,Klinkhamer:2007pe,
Thomas:2009uh,Maggiore:2010wr} (for a detailed pedagogical discussion, see~\cite{Sola:2013gha}). The presence of two distinct fundamental scales emerges as a direct
consequence of the auxiliary conditions on the physical Hilbert space that are required by the unitarity condition of an underlined quantum theory.
Remarkably, the Veneziano ghost effect emerges as a positive contribution to the vacuum energy density with a time-dependent equation of state,
which can be considered as a source of DE.

Very recently, many features of the early-time acceleration were explained within a strongly-coupled QCD-like theory (denoted as $\overline{\rm QCD}$)
in terms of an auxiliary topological field in~\cite{Zhitnitsky:2013pna,Zhitnitsky:2014aja,Zhitnitsky:2015dia}. Indeed, the de-Sitter phase
can be dynamically initiated in the expanding Universe by a topological (auxiliary) non-propagating field, which does not possess a canonical kinetic
term in analogy to the topologically-ordered phases in condensed matter systems. As a characteristic postulate in this approach,
the vacuum energy in the FLRW Universe is expected to behave linearly with the Hubble parameter, {\it i.e.}:
\begin{eqnarray}
\label{QCD-vac}
\Delta \epsilon_{\rm vac} = H\,\Lambda_{\overline{\rm QCD}}^3 + h.o. \,, \qquad
H \simeq \frac{\kappa}{3}\Lambda_{\overline{\rm QCD}}^3 \ll \Lambda_{\overline{\rm QCD}} \,,
\end{eqnarray}
in full consistency with the Zeldovich relation, where $\Lambda_{\overline{\rm QCD}}$ is the energy scale of the QCD-like theory.
The above postulate has been confronted with observational data in, e.g.,~\cite{Cai:2010uf,Sheykhi:2011xz,Sheykhi:2011nb,
RozasFernandez:2011je,Feng:2012wx,Cai:2012fq,Feng:2012gr,Alcaniz:2012mh,Malekjani:2012wc}, showing the consistency of
the linear scaling with available data, in tension with~\cite{GarciaSalcedo:2013xs}, claiming that no critical point associated
with matter dominance is found in the physical phase space of the model. Anyway, the scaling (\ref{QCD-vac}) is in variance to the conventionally
accepted scaling behavior $\Delta \epsilon_{\rm vac}\sim H^2$ based on the principles of locality and general covariance \cite{Shapiro:1999zt,Shapiro:2000dz}
due to the fact that in strongly-coupled field theories, the locality is violated, at least in the Minkowski background. The topological vacuum
energy in $\overline{\rm QCD}$ has a non-dispersive nature and cannot be understood in terms of any local propagating degrees of freedom.
Similar arguments can be applied for understanding the QCD origin of DE at the late-time acceleration epoch \cite{Holdom:2010ak}. Such promising
developments of the inflaton and DE interpretations in strongly-coupled gauge theories clearly provide a strong motivation for further investigations in this direction.

\section{Graviton-Exchange Correction to the QCD Ground State: A Pedagogical Outlook}

In order to illustrate the rigorous procedure of the QCD-induced CC computation in the conventional QFT framework, consider semiclassical gravity
coupled to the quantum fluctuations in the non-perturbative QCD vacuum \cite{Pasechnik:2013poa}. The metric operator $\hat{g}^{\mu\nu}$
contains the $c$-number part $g^{\mu\nu}$, the macroscopic space-time metric in which all the covariant derivatives and lowering/raising index
operations are defined (in this section, we adopt the standard GR notations, e.g., the covariant differentiation w.r.t. $x^\mu$ is denoted
by a semicolon, while an ordinary derivative by a comma, {\it etc.}), and operator part $\Phi_\mu^\nu$, the quantum graviton field satisfying \cite{QG-1,QG-2}:
\begin{equation}
\label{Phiav}
\langle 0|\Phi_\mu^\nu |0\rangle = 0\,,
\end{equation}
where the averaging is performed over the Heisenberg state vector $|0\rangle$. The corresponding action~reads:
\begin{equation}
\label{gravS}
 S = \int{Ld^4x}\,, \hspace{5mm} L=-\frac{1}{2\kappa}
 \sqrt{-\hat{g}}\hat{g}^{\mu\nu}\hat{R}_{\mu\nu} + {\cal L}_{\rm QCD}\,,
\end{equation}
where $\hat{R}_{\mu\nu}$ are the curvature operator and ${\cal L}_{\rm QCD}$ is the properly generalized
Lagrangian density of QCD. Varying w.r.t. the macroscopic metric (or graviton field) leads to the standard operator
equations of motion of semiclassical gravity. Then, averaging over the Heisenberg state vector, one finally
arrives at the Einstein equations for $g^{\mu\nu}$:
\begin{equation}
\displaystyle \frac{1}{\kappa}\left(
 R_\mu^\nu -\frac{1}{2}\delta_\mu^\nu R
 \right)=\langle 0 |\hat{T}_\mu^\nu | 0 \rangle\,,
\label{E}
\end{equation}
and the equations for the graviton field:
\begin{equation}\label{graveq2}
\begin{array}{c}
    \psi_{\mu;\rho}^{\nu;\rho}
    -\psi_{\mu;\rho}^{\rho;\nu}
    -\psi_{\rho;\mu}^{\nu;\rho}
    +\delta_\mu^\nu\psi_{\rho;\sigma}^{\sigma;\rho}
    +\psi_\mu^\rho R_\rho^\nu
    +\psi_\rho^\nu R_\mu^\rho
    -\delta_\mu^\nu R_\rho^\sigma \psi_\sigma^\rho =
    2\kappa\left(\hat{T}_\mu^\nu - \langle 0 |\hat{T}_\mu^\nu | 0 \rangle\right)\,.
\end{array}
\end{equation}
Here,
$$
 \psi_\mu^\nu = \Phi_\mu^\nu - \frac{1}{2}\delta_\mu^\nu \Phi\,,
$$
and the total operator of the energy-momentum tensor:
\begin{equation}\label{Tandgrav}
 \hat{T}_\mu^\nu=\hat{T}_{\mu(G)}^\nu + \frac{1}{2}
 \left(\delta_\rho^\nu \delta_\mu^\sigma + g^{\nu \sigma}g_{\mu \rho}\right)
 \left(\frac{\hat{g}}{g}\right)^{1/2}
 \hat{g}^{\rho \tau}\hat{T}^{\rm QCD}_{\tau\sigma}
\end{equation}
accounting for the graviton field contribution:
\begin{equation}
\label{Tgrav}
\begin{array}{c}
\displaystyle
 \hat{T}_{\mu (G)}^\nu = \frac{1}{4\kappa}
 \left(
   \psi_{\sigma;\mu}^\rho \psi_\rho^{\sigma;\nu}
   -\frac{1}{2}\psi_{;\mu}\psi^{;\nu}
   -\psi_{\mu;\sigma}^\rho \psi_\rho^{\sigma;\nu} -\psi_\rho^{\nu;\sigma} \psi_{\sigma;\mu}^\rho\right)
\\[3mm] \displaystyle
 -\frac{1}{8\kappa}\delta_\mu^\nu
 \left(
   \psi_{\sigma;\gamma}^\rho \psi_\rho^{\sigma;\gamma}
   -\frac{1}{2}\psi_{;\gamma} \psi^{;\gamma}
   -2\psi^\rho_{\gamma;\sigma} \psi_\rho^{\sigma;\gamma}
 \right)
\\[3mm] \displaystyle
 -\frac{1}{4\kappa}
 \left(
   2\psi_\gamma^\rho \psi_\mu^{\nu;\gamma}
   -\psi_\gamma^\nu \psi_\mu^{\rho;\gamma}
   -\psi_\mu^\gamma \psi_{\hspace{2mm};\gamma}^{\nu \rho}
    +\psi_\mu^{\gamma;\nu} \psi_\gamma^\rho
   +\psi_{\gamma;\mu}^{\nu} \psi^{\gamma\rho} +
   \delta_\mu^\nu \left(\psi_\sigma^\gamma \psi_\gamma^\rho\right)^{;\sigma} \right)_{;\rho}
\\[3mm] \displaystyle
-\frac{1}{4\kappa}
 \left(
   \psi_\mu^\sigma \psi_\gamma^\rho R_\rho^\nu
   +\psi_\gamma^\nu \psi_\rho^\gamma R_\mu^\rho
   -\delta_\mu^\nu \psi_\rho^\gamma \psi_\gamma^\sigma R_\sigma^\rho
 \right)
 +O(\psi^3)\,,
\end{array}
\end{equation}
and the effective operator energy-momentum tensor of QCD $\hat{T}^{\rm QCD}_{\rho \sigma}$.
The latter takes the form of Equation~(\ref{S-tem}) for pure gluodynamics with a trace anomaly, which
is sufficient for illustration purposes here. Introducing a characteristic scale of non-perturbative QCD
fluctuations in terms of the effective correlation length of condensate fluctuations $L_{g}^{-2}\simeq (1.5 \;\text{GeV})^2$
as follows:
\[
\displaystyle
\ln\frac{e\langle0|J|0\rangle}{(\xi \lambda)^4}\to 4\ln\frac{L_{g}^{-1}}{\Lambda_{\rm
QCD}} \,,
\]
and keeping the average of the trace anomaly only, one arrives at the following approximated
expression for the QCD energy-momentum tensor in the Minkowski background:
\vspace{-6pt}
\begin{equation}
\begin{array}{c}
\displaystyle T_{\mu}^{\nu,{\rm QCD}}\simeq \frac{b\alpha_s}{2\pi}\left(-F^a_{\mu \rho}F_a^{\nu \rho} + \frac14\delta_\mu^\nu
F^a_{\sigma\rho}F_a^{\sigma\rho}\right) \ln\frac{L_{g}^{-1}}{\Lambda_{\rm QCD}}
-\delta_\mu^\nu\frac{b}{32}\langle 0 |\frac{\alpha_s}{\pi}F^a_{\sigma\rho} F_a^{\sigma\rho}| 0 \rangle\,.
\label{S-tem0}
\end{array}
\end{equation}
with the gluon field equations of motion in the form:
\begin{equation}
\displaystyle D^{ab}_\nu F^{\mu\nu}_b=0\,, \qquad
D^{ab}_\nu = \delta^{ab}\partial_\nu - g_s f^{abc}A_\nu^c\,.
\label{qcd-0}
\end{equation}
These expressions should then be generalized to an operator covariant form.
For example, Equation~(\ref{qcd-0}) reads:
\begin{equation}
\displaystyle
\left(\delta^{ab}\frac{\partial}{\partial
x^\nu}-g_s f^{abc}\hat{A}_\nu^c\right)\sqrt{-\hat{g}}\hat{g}^{\mu \rho}\hat{g}^{\nu \sigma}\hat{F}^b_{\rho \sigma}=0\,.
\label{qcd-cov}
\end{equation}

Next, let us estimate the contribution of graviton-exchange interactions to the QCD ground state energy
accounting for the first-order (linear) non-vanishing terms in gravitational constant $G$, such as $O(\alpha_sG)$
terms to the one-loop QCD approximation. Due to a smallness of the typical QCD space-time scales compared to
the cosmological scales in the modern Universe, the induced quantum fluctuations of the metric should
be considered at the Minkowski background. Then, the trace of the macroscopic Einstein equations (\ref{E}):
\[
R+4\kappa\, \epsilon^{\rm QCD}_{\Lambda} = 0
\]
gives rise to the QCD-induced correction to the $\Lambda$-term density:
\begin{equation}
 \epsilon^{\rm QCD}_{\Lambda} = -\frac{b}{32}
 \langle 0 |\frac{\alpha_s}{\pi} \left(\frac{\hat{g}}{g}\right)^{1/2}
 \hat{g}^{\mu \rho} \hat{g}^{\nu \sigma}\hat{F}^a_{\mu\nu}\hat{F}^a_{\rho \sigma}|0\rangle+
 \frac14\langle 0|\hat{T}_{(G)}|0\rangle\,,
\label{E1}
\end{equation}
where the last term contains the QCD-induced graviton component. Applying the gluon and graviton field
Equations (\ref{qcd-cov}) and (\ref{graveq2}) accounting for the QCD and graviton energy-momentum tensors (\ref{S-tem0})
and (\ref{Tgrav}), respectively, it is straightforward to show that the resulting contribution to the $\Lambda$-term density
takes a remarkably simple form convenient for phenomenological analysis \cite{Pasechnik:2013poa}:
\begin{equation}
\displaystyle \epsilon^{\rm QCD}_{\Lambda} =
 -\frac{b}{16}\ln\frac{L_{g}^{-1}}{e\Lambda_{\rm QCD}}\,\langle 0|
 \frac{\alpha_s}{\pi} F^a_{\mu \rho }F_a^{\nu \rho }\left(\psi_\nu^\mu -
 \frac{1}{4}\delta_\nu^\mu \psi\right)|0\rangle\,,
 \label{laint1}
\end{equation}
where the graviton field induced by non-perturbative QCD vacuum fluctuations is to be found as a solution
of the corresponding equation of motion:
\begin{equation}
\begin{array}{c}
 \displaystyle
    \psi_{\mu,\,\rho}^{\nu,\,\rho}
    -\psi_{\mu,\,\rho}^{\rho,\,\nu}
    -\psi_{\rho,\,\mu}^{\nu,\,\rho}
    +\delta_\mu^\nu\psi_{\rho,\,\sigma}^{\sigma,\,\rho}
 =\frac{\kappa\alpha_s\, b}{\pi}\left(-F^a_{\mu \rho}F_a^{\nu \rho}+\frac14\delta_\mu^\nu
F^a_{\sigma\rho}F_a^{\sigma\rho}\right)\ln\frac{L_g^{-1}}{\Lambda_{\rm QCD}}\,.
\label{fluc}
\end{array}
\end{equation}
The latter is most conveniently performed in the Fock gauge $\psi_{\mu;\,\nu}^\nu=0$ (for more details, see~\cite{Pasechnik:2013poa}).
Finally, applying the compensation condition for the correlation functions (\ref{dcomp}), one arrives at the following expression
consistent with the Zeldovich relation and, hence, with the linear scaling (\ref{QCD-vac}):
\begin{equation}
\begin{array}{c}
\displaystyle \epsilon^{\rm QCD}_{\Lambda} \simeq -\pi G\langle
0|:\frac{\alpha_s}{\pi}F_{\mu\nu}^aF_a^{\mu\nu}:|0\rangle^2\times
\left(\frac{b}{8}\right)^2
\ln\frac{L_{g}^{-1}}{e\Lambda_{\rm
QCD}}\ln\frac{L_{g}^{-1}}{\Lambda_{\rm QCD}} \int
d^4y\mathcal{G}(y)D^2(y) \simeq
\\[5mm]
\displaystyle \simeq \Delta\,(1\pm0.5)\times 10^{-41} \;\text{GeV}^4 \,,
\end{array}
\label{fin}
\end{equation}
where $\langle 0|:\frac{\alpha_s}{\pi}F_{\mu\nu}^aF_a^{\mu\nu}:|0\rangle$ is the phenomenologically-constrained (local) gluon
condensate (\ref{top1}) and $\Delta$ is the positively-definite dimensionless parameter:
\begin{equation}
\displaystyle \Delta = -\frac{1}{L_g^2}\int
d^4y\mathcal{G}(y)D^2(y) > 0\,, \label{Del}
\end{equation}
defined in Euclidean four-space in terms of the correlation function $D=D(x)$ (\ref{dcomp}) and the Green function
$\mathcal{G}=\mathcal{G}(x)$ satisfying the Green equation $\mathcal{G}_{,\,l}^{,\,l}=-\delta(x-x')$.
The parameter $\Delta$ is thus determined by unknown dynamics of the non-perturbative QCD vacuum in Minkowski
spacetime and should to be established, e.g., in effective field theory approaches or in lattice QCD. Note, the inclusion of
non-perturbative light quark fluctuations effectively changes the one-loop $\beta$-function coefficient only and, thus,
does not strongly affect the overall estimate (\ref{fin}). Note, the result (\ref{fin}) is based on the standard approach
to weak semiclassical gravity and effective low-energy QCD with the trace anomaly and does not incorporate any strong
physical assumptions or ideas beyond the standard QFT.

Unfortunately, one cannot provide an accurate estimate for the $\Delta$ parameter at the current level of theoretical
understanding of the non-perturbative QCD vacuum. Individually, the terms $D_{(-)}(x-x')$ and $D_{(+)}(x-x')$ in
the complete correlation function $D(x-x')$ lead to $\Delta \sim 1$. However, their difference may provide a small,
but non-zeroth $\Delta$, due to a shift in scales triggered by the chiral symmetry breaking in low energy QCD.
In terms of small current quark $m_u,\,m_d,\,m_s$ masses as naturally small QCD parameters responsible for the
chiral symmetry breaking, an induced mismatch between the characteristic scales of $D_{(-)}$ and $D_{(+)}$ fluctuations:
\[
\displaystyle 1/L_{+}\sim 1/L_{-}\sim 1/L_g\,, \qquad
|1/L_{+}-1/L_{-}| \sim m_u+m_d+m_s \sim 0.1\, {\rm GeV}
\]
leads to an order-of-magnitude estimate:
\begin{equation}
\displaystyle \Delta = k\cdot \frac{(m_u+m_d+m_s)^2L_g^2}{(2\pi)^4}>0\,, \qquad k \sim 1 \,, \qquad
\epsilon^{\rm QCD}_{\Lambda} \sim \Lambda_{\rm cosm} \sim 10^{-47}\,{\rm GeV}^4 \,.
\label{Del1}
\end{equation}
Therefore, this naive, but phenomenologically-motivated estimate provides the value for the QCD-induced $\Lambda$-term
density of the same order of magnitude as the observed CC value (\ref{Lcosm}) with the correct sign. The above simple
illustration independently validates the fact elaborated in~\cite{Zhitnitsky:2013pna,Zhitnitsky:2014aja,Zhitnitsky:2015dia}
that in a strongly-coupled field theory, such as QCD, the principle of locality is violated, leading to an unexpected linear
dependence of the topological DE density on the Hubble scale, in consistency with Zeldovich scaling relation and
the observed CC. The above approach, however, misses important, but yet inaccessible information on the real-time
dynamics of the QCD vacuum, which prevents a more rigorous analysis of the dynamics of the QCD-induced DE in the expanding Universe.

\section{Summary}

Yet, there is no common consensus in the theory community on what the resolution for the CC problem should be.
In this short review, I provided a quick outlook of the most promising DE/CC interpretations and theoretical developments
that can be found within conventional QFT and particle physics concerning, in particular, YM theories with a non-trivial ground state.

Remarkably, the quantum effects strongly affect the dynamics of the gluon condensate at cosmological times, switching it from a relativistic matter
$p_{\rm YM}=\varepsilon_{\rm YM}/3$ in the pure classical case (see Section~\ref{Sec:YM-Cosm}) to either vacuum-type CC (without subtraction (\ref{presc-1})
imposed) or to a dust-type matter. This is due to the fact that the equations of motion are unstable w.r.t. quantum fluctuations, substantially affecting
the ground state properties, even at cosmologically large time scales via the conformal~anomaly.

In many studies throughout the literature, it has been shown that strongly-coupled gauge theories, such as QCD, have
all of the features required to describe the basic characteristics of both the late (DE) and early (inflation) time acceleration
epochs in the Universe's evolution. These features are studied within two distinct approaches based on specific properties of
the topological non-perturbative QCD vacuum:

(i) The gluon condensate in the expanding Universe has two components contributing to the QCD ground state with opposite signs
which asymptotically (in the IR limit of the theory) reach attractor states with energy densities exactly eliminating each other
at $t\to\infty$. The net QCD-induced DE component can then be understood as an uncompensated positive remnant of such
a gross~cancellation.

(ii) Alternatively, the energy density of the gluon condensate in the Minkowski background should be subtracted from that in the FLRW
background of the expanded Universe, leading to a non-zeroth remnant, which is identified with the observed CC $\Lambda_{\rm cosm}$.
As a characteristic prediction of the non-local topological QCD vacuum evolution in the expanding Universe, the CC in this (de-Sitter) case
is expected to scale linearly with the Hubble parameter in consistency with the Zeldovich scaling relation $\Lambda_{\rm cosm}\propto
\Lambda_{\rm QCD}^6/M_{\rm PL}^2$. A similar result naturally comes out in the framework of quasiclassical gravity where the $\Lambda$-term
emerges as a leading-order gravitational correction to the QCD ground-state energy density induced by the graviton-exchange interactions
in the QCD~vacuum.

Both approaches rely on an elimination of the microscopic (in particular, QCD) vacua contributions to the ground state energy of the Universe,
either dynamically or by a phenomenologically reasonable subtraction condition. Clearly, there is a long way to go towards a complete
dynamical understanding of the DE, as well as the dynamical properties of the topological vacuum in strongly-coupled (QCD-like) field theories.
However, one could share a careful optimism that a link between those has now been established, which certainly requires further deeper studies.

\vspace{12pt}
\noindent{\bf Acknowledgments:}
This work was supported by the Swedish Research Council, contract Number 621-2013-4287.

\conflictofinterests{\textbf{Conflicts of Interest:} The author declares no conflict of interest.}

\bibliographystyle{mdpi}
\renewcommand\bibname{References}

\end{document}